\documentclass[9pt,twocolumn,twoside]{osajnl}

\journal{ao} 

\setboolean{shortarticle}{false}
\usepackage{graphicx}
\usepackage{lineno}
\usepackage{url}
\usepackage{booktabs, multirow,tabularx}
\newcolumntype{Y}{>{\arraybackslash}X}
\usepackage{amsmath}
\DeclareMathOperator{\sinc}{sinc}
\usepackage{xcolor}
\AfterPreamble{\hypersetup{
  urlcolor=blue,
}}
\title{The Simons Observatory: Characterizing the Large Aperture Telescope Receiver with Radio Holography}

\author[1,*]{Grace E. Chesmore}
\author[2]{Kathleen Harrington}
\author[1]{Carlos E. Sierra}
\author[3]{Patricio A. Gallardo}
\author[1]{Shreya Sutariya}
\author[1]{Tommy Alford}
\author[4]{Alexandre E. Adler}
\author[5]{Tanay Bhandarkar}
\author[6]{Gabriele Coppi}
\author[4]{Nadia Dachlythra}
\author[1]{Joseph Golec}
\author[4]{Jon Gudmundsson}
\author[5]{Saianeesh K. Haridas}
\author[7]{Bradley R. Johnson}
\author[5]{Anna M. Kofman}
\author[5]{Jeffrey Iuliano}
\author[1,2,8-10]{Jeff McMahon}

\author[11]{Michael D. Niemack}
\author[5]{John Orlowski-Scherer}
\author[5]{Karen Perez Sarmiento}
\author[13]{Roberto Puddu}
\author[12]{Max Silva-Feaver}
\author[10]{Sara M. Simon}
\author[5]{Julia Robe}
\author[14]{Edward J. Wollack}
\author[15]{Zhilei Xu}

\affil[1]{Department of Physics, University of Chicago, 5720 South Ellis Avenue, Chicago, IL 60637, USA}
\affil[2]{Department of Astronomy and Astrophysics, University of Chicago, 5640 S. Ellis Ave., Chicago, IL 60637, USA}
\affil[3]{Kavli Institute for Cosmological Physics, University of Chicago, 5640 S. Ellis Ave., Chicago, IL 60637, USA}
\affil[4]{The Oskar Klein Centre, Department of Physics, Stockholm University, SE-106 91 Stockholm, Sweden}
\affil[5]{University of Pennsylvania, Philadelphia, PA 19104, USA}
\affil[6]{Department of Physics, University of Milano-Bicocca, Piazza della Scienza 3, 20126 Milano, Italy; National Institute for Nuclear Physics (INFN), Sezione di Milano-Bicocca, Piazza della Scienza 3, 20126 Milano, Italy}
\affil[7]{University of Virginia, Department of Astronomy, Charlottesville, VA 22904}
\affil[8]{Kavli Institute for Cosmological Physics, University of Chicago, 5640 S. Ellis Ave., Chicago, IL 60637, USA}
\affil[9]{Enrico Fermi Institute, University of Chicago, Chicago, IL 60637, USA}
\affil[10]{Fermi National Accelerator Laboratory, Batavia, IL, USA}
\affil[11]{Department of Physics, Cornell University, Ithaca, NY 14853 USA; Department of Astronomy, Cornell University, Ithaca, NY 14853 USA}
\affil[12]{University of California, San Diego Department of Physics 9500 Gilman Dr. La Jolla, CA 92093}
\affil[13]{Centro de Astro-Ingeniería, Facultad de Ingeniería, Pontificia Universidad Católica de Chile, Av. Vicuña Mackenna 4860, 7820436 Macul, Santiago, Chile}
\affil[14]{NASA Goddard Space Flight Center, 8800 Greenbelt Rd, Greenbelt, MD 20771, USA}
\affil[15]{MIT Kavli Institute, Massachusetts Institute of Technology, 77 Massachusetts Avenue, Cambridge, MA 02139, USA}

\affil[*]{Corresponding author: chesmore@uchicago.edu}

\begin{abstract}

We present near-field radio holography measurements of the Simons Observatory Large Aperture Telescope Receiver optics.  These measurements demonstrate that radio holography of complex millimeter-wave optical systems comprising cryogenic lenses, filters, and feed horns can provide detailed characterization of wave propagation before deployment.  We used the measured amplitude and phase, at 4\,K, of the receiver near-field beam pattern to predict two key performance parameters: 1) the amount of scattered light that will spill past the telescope to 300\,K and 2) the beam pattern expected from the receiver when fielded on the telescope.  These cryogenic measurements informed the removal of a filter, which led to improved optical efficiency and reduced side-lobes at the exit of the receiver.  Holography measurements of this system suggest that the spilled power past the telescope mirrors will be less than 1\% and the main beam with its near side-lobes are consistent with the nominal telescope design.  This is the first time such parameters have been confirmed in the lab prior to deployment of a new receiver.   This approach is broadly applicable to millimeter and sub-millimeter instruments.

\end{abstract}

\setboolean{displaycopyright}{true}
\begin{document}
\maketitle
\section{Introduction}
Simons Observatory (SO) will observe the cosmic microwave background (CMB) temperature and polarization signals using multiple millimeter-wave telescopes~\cite{gali18, so19}.  One Large Aperture Telescope (LAT)~\cite{Niemack:16, Gudmundsson:21,Parshley_2018} and three Small Aperture Telescopes (SAT)~\cite{ali20} together will measure the CMB anisotropies.  SO will provide new constraints on inflationary signals, neutrino mass, and particles beyond the standard model while further improving our understanding of dark energy and galaxy evolution and the era of cosmic reionization~\citep{so19}. 
\begin{figure}
    \centering
    \includegraphics[width = .47\textwidth]{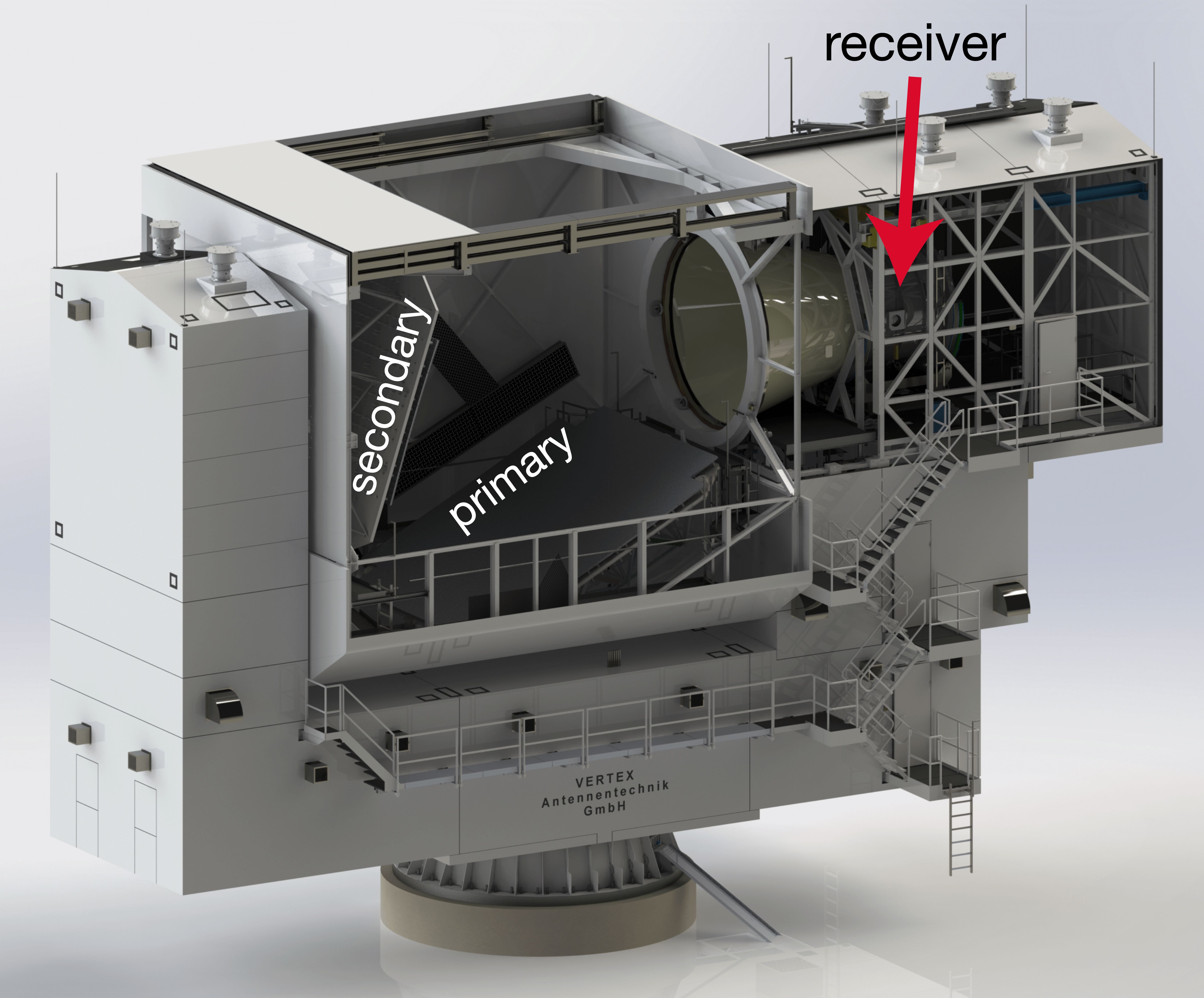}
    \caption{The Simons Observatory (SO) Large Aperture Telescope (LAT), featuring a segmented primary and secondary mirror.  The focus of the mirror is hidden inside the conical baffle near the front of the receiver.  The receiver can hold up to thirteen optics tubes.}
    \label{fig:lat}
\end{figure}
The LAT is a crossed-Dragone telescope~\cite{6773968,Niemack:16,2021RNAAS...5..100X} developed in collaboration with the Fred Young Sub-millimeter Telescope (FYST)~\cite{ccat,aravena2019ccatprime} Collaboration.  The LAT Receiver (LATR) can hold up to 13 optics tubes, which can accommodate more than 60,000 detectors distributed across 39 detector wafers for CMB studies~\cite{Parshley_2018,2021ApJS25623Z,mccarrick2021simons}.  Each optics tube holds a set of lenses, filters, and baffles which couple light from the telescope onto a set of three polarization sensitive detector arrays.   Realizing the SO science goals depends on achieving stringent sensitivity requirements and controlling systematic effects to be subdominant to the statistical noise.   To achieve this, the optics tubes must have clean beams with well-controlled side-lobe power.  

Testing these aspects of the optics tube performance in the lab is challenging.  Previously, these parameters were verified on ground-based CMB radio telescope systems upon deployment~\cite{alma_holog,2007A&A...465..679N}.  In this paper, we present a new approach to laboratory testing of these systems using the technique of near-field radio holography.  Holography is an instrumentation technique used to measure the complex monochromatic electric field wavefront using the interference between a modulated and reference signal.  Radio holography takes advantage of the antenna theory relationship: the far-field radiation pattern of a reflector antenna is the Fourier Transformation of the field distribution in the aperture plane of the antenna~\cite{alma_holog}.

Within radio beam characterization, several techniques exist; 1) scalar beam pattern characterization~\cite{doi:10.1063/1.3292308}, or measuring only the magnitude of the wavefront, 2) vector beam pattern characterization~\cite{2020JLTP..199..156Y,7740846}, or measuring the amplitude and phase of the wavefront, and 3) holographic beam pattern measurement~\cite{387181,7740846}, which we explore here.  While scalar beam pattern characterization requires the beam to be measured at multiple planes along the propagation axis, and vector beam characterization records one complex map, holography records two beam maps (one modulated by the optical element and the other used as a reference) to reconstruct the complex wavefront~\cite{4584681,alma_holog}.  Additional use of radio beam characterization is phase retrieval holography, with applications in electron microscopy, optical imaging, and crystallography~\cite{7078985}.

Near-field holography allows us to study the wavefront as it emerges from the optics tube, based on the reciprocal theorem, from the cryostat.  Using Fresnel diffraction (FD)~\cite{Goodman2005-ne}, these measured fields can be propagated through the optical system to determine the spilled power past the mirrors of the telescope and the far-field beam pattern of the telescope fed by this receiver.  Moreover, these beams are useful for the identification and mitigation of optical problems within the receiver (i.e. optical aberrations, focus, scattered power, etc.).  These measurements enable a detailed verification of system-level optical performance prior to the deployment of a receiver on a telescope.

In Section~\ref{sec:optics_tube} we describe the optical design and components of the SO optics tube.  In Section~\ref{sec:meas_method} we describe the measurement approach including the cryogenic receiver (\ref{sec:meas_method}.\ref{sec:cryo_rec}) and holography hardware (\ref{sec:meas_method}.\ref{sec:meas_hardware}) required for measuring beam maps (full details can be found in Appendix~\ref{sec:appendix_hardware}).  In Section~\ref{sec:results} we present the measured beam maps.   Section~\ref{sec:prop_fields} discusses analysis methods to propagate the measured beams into the far-field using FD.  We present characterization of the optics tube with and without an infrared-blocking filter in Section~\ref{sec:filter}.  Section~\ref{sec:code} details the publicly available code.  We conclude with a discussion of future applications of this approach in Section~\ref{sec:discussion}.  
\begin{figure}[H]
    \centering
    \includegraphics[width = .47\textwidth]{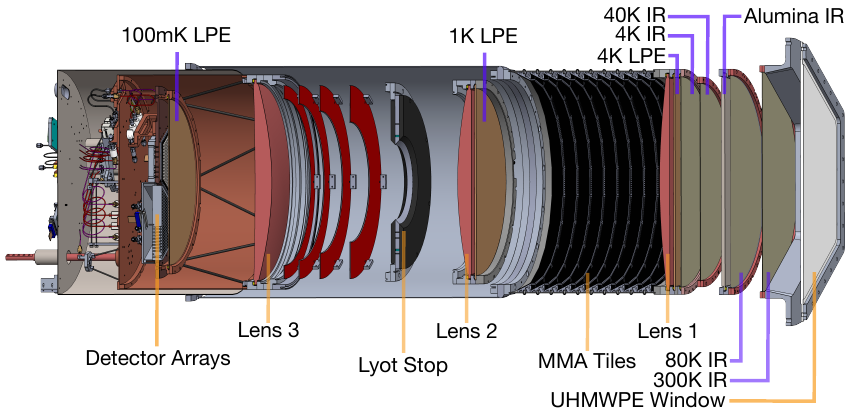}
    \caption{The Simons Observatory LATR optics tube.  Light enters the optics tube from the right through an ultra-high-molecular-weight polyethylene (UHMWPE) window, and travels through a series of Infrared-Blocking filters (IR), including one alumina IR filter, and passes through the 4\,K Low-Pass Edge mesh filter (LPE). The light is focused on the detector arrays on the left by the three lenses, with two additional LPE filters near the 1\,K and 100\,mK stages.  The components within the optics tube are further described in~\cite{xu/etal:2020c}.}
    \label{fig:latrt}
\end{figure}
\section{SO Large Aperture Telescope Optics Tubes Design}
\label{sec:optics_tube}
\begin{figure}[ht]
    \centering
    \includegraphics[width = .47\textwidth]{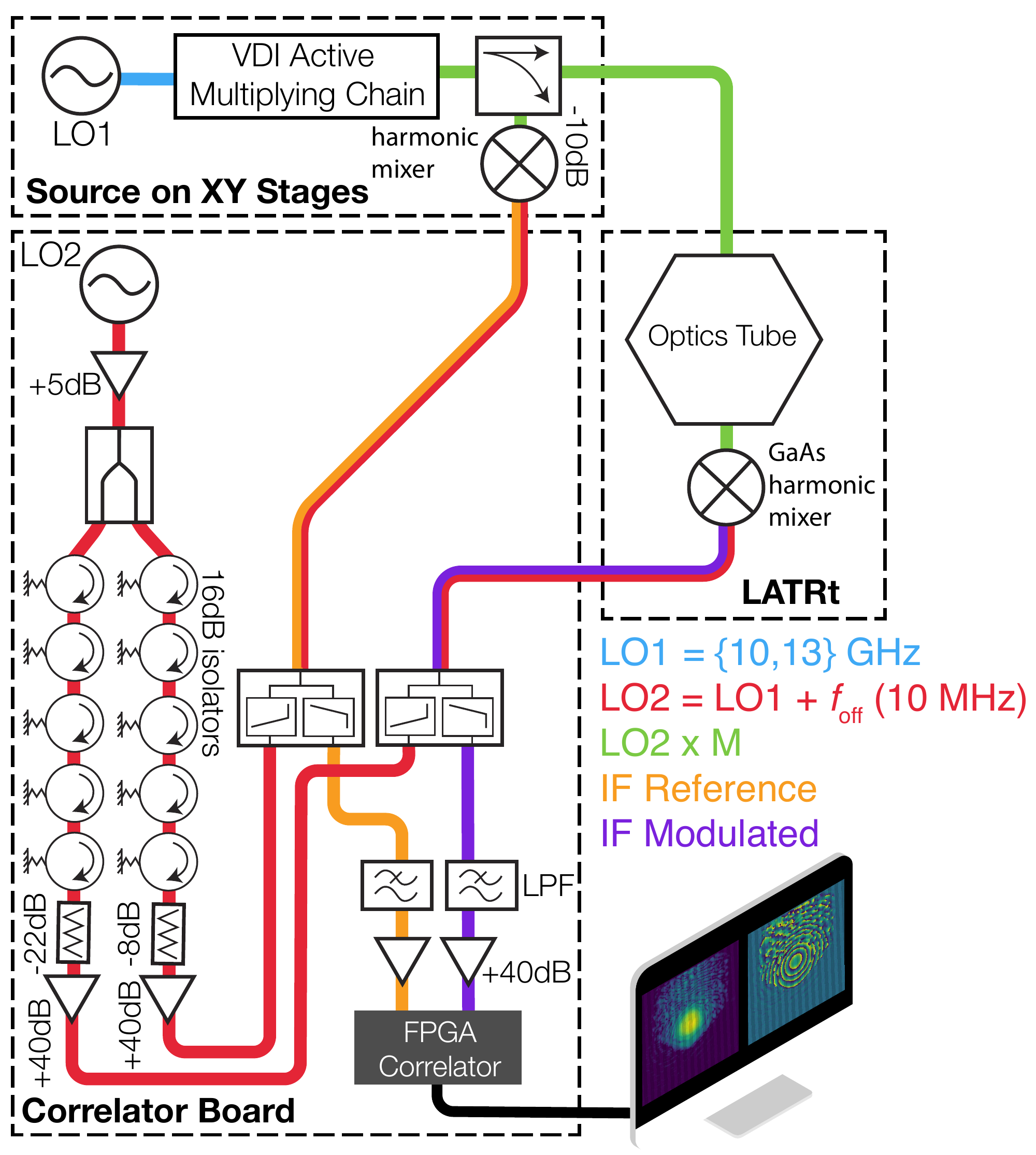}
    \caption{Schematic of holography setup.  Two local oscillators (LOs) supply frequencies (LO1 and $\text{LO2} = \text{LO1} + f_{\text{off}}$), one of which is multiplied by the active multiplier factor $M$ , while the other serves to down convert the reference and modulated signals into the intermediate frequency.}
    \label{fig:setup}
\end{figure}
The LAT is a crossed-Dragone telescope with a 6\,m pupil diameter (Fig.~\ref{fig:lat}).  The LAT Receiver is designed to house thirteen optics tubes, which guide photons onto cryogenic detectors.  These optics tubes must maintain excellent beam quality while also limiting the wide-angle scattering.  The scattering is critical to the ultimate sensitivity of the SO LAT since every percent that is scattered leads to 3\,K of extra noise and a 15(10)\,\% reduction in mapping speed in the 90(150)\,GHz bands which are critical for the SO science.

Figure~\ref{fig:latrt} shows the LATR optics tube~\cite{xu/etal:2020c}.  Light enters through a 3\,mm thick ultra-high molecular weight polyethylene hexagonal window with an anti-reflection coating~\cite{zhu18}.  Three anti-reflection coated silicon lenses~\cite{Datta:13,golec20} control the beam size and shape, which re-images from the focal plane onto three hexagonal detector wafers.  Each lens is accompanied by a low-pass edge filter (LPE)~\cite{10.1117/12.673162}.  The light is coupled onto the detector wafers using arrays of drilled spline profile feed horns and the polarization is coupled onto the wafer with an orthomode transducer~\cite{10.1117/12.2313405,2022arXiv220104507H}.  Along this path, the light passes through a succession of band-defining and infra-red blocking filters~\cite{10.1117/12.673162}.  From the sky side these are a 300\,K IR blocking filter, an 80\,K IR blocking filter, an 80\,K IR rejecting metamaterial anti-reflection coated alumina filter~\cite{golec20,golec2022}, and a 40\,K IR blocking filter~\cite{10.1117/12.2561720} prior to entering the optics tube.  The filter configuration is described in~\cite{zhu18}.  Between lenses 1 and 2, the walls of the optics tube are coated with baffling metamaterial tiles, described in~\cite{Xu:21}, which control scattering.  The full cold-optical design is described in~\cite{dicker2019cold}.  All optical elements in the optics tube are between 4\,K and 100\,mK. 
\begin{figure}[ht!]
    \centering
    \includegraphics[width = .49\textwidth]{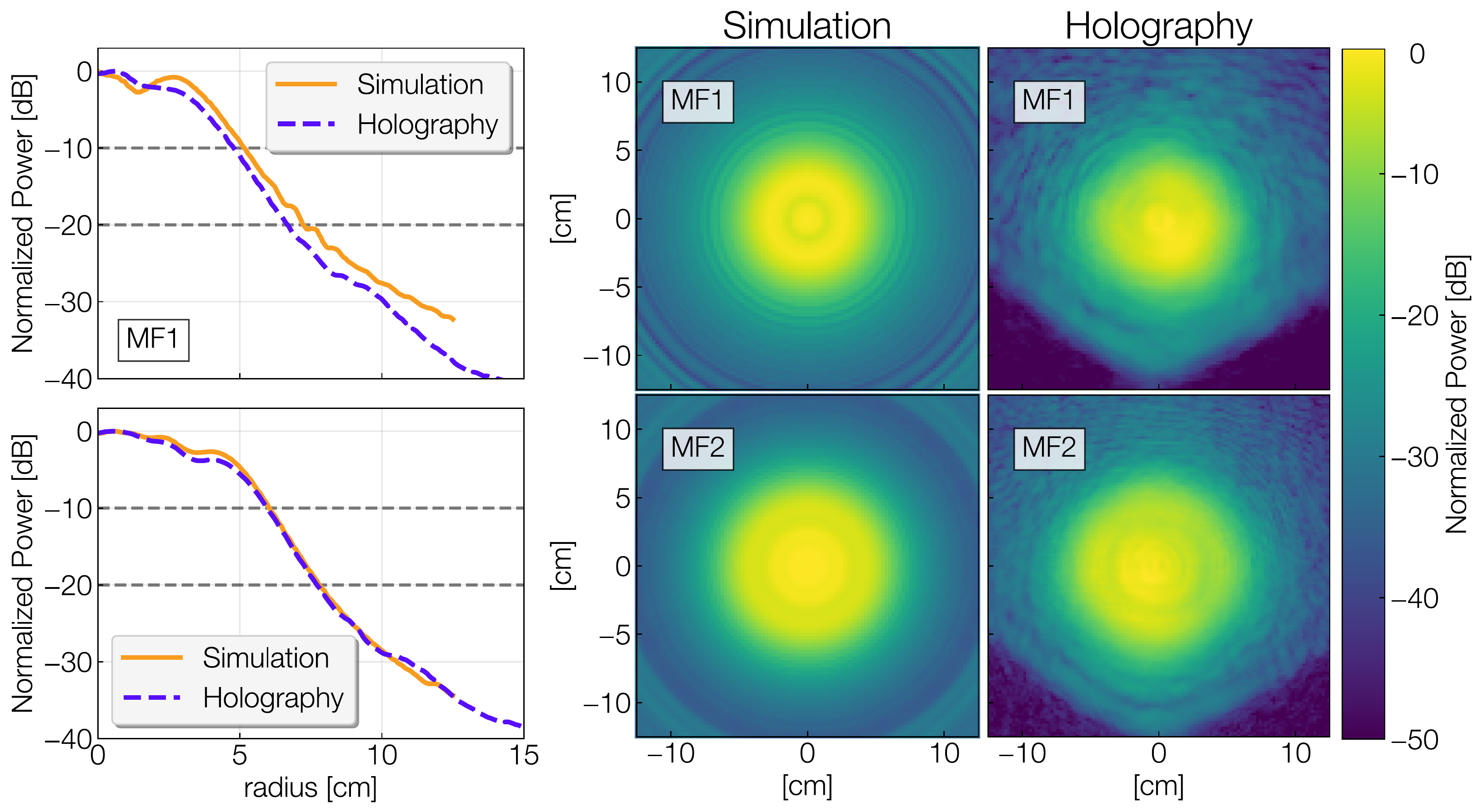}
    \caption{Near-field band-averaged MF1 and MF2 beams, simulated and measured with radio holography. Left panel: Radially binned.  Right panel: 2D simulated and measured near-field beams.}
    \label{fig:nearfields}
\end{figure}
\begin{figure*}[ht]
    \centering
    \includegraphics[width = .97\textwidth]{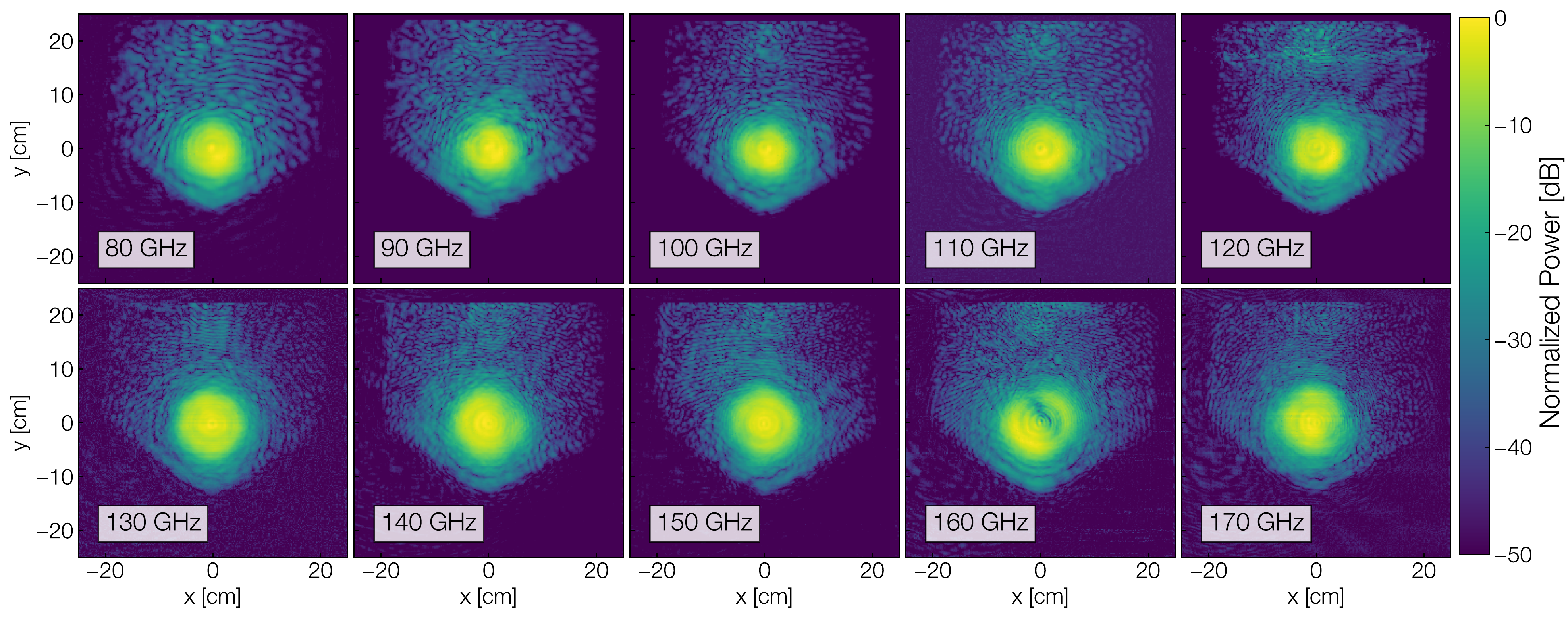}
    \includegraphics[width = .97\textwidth]{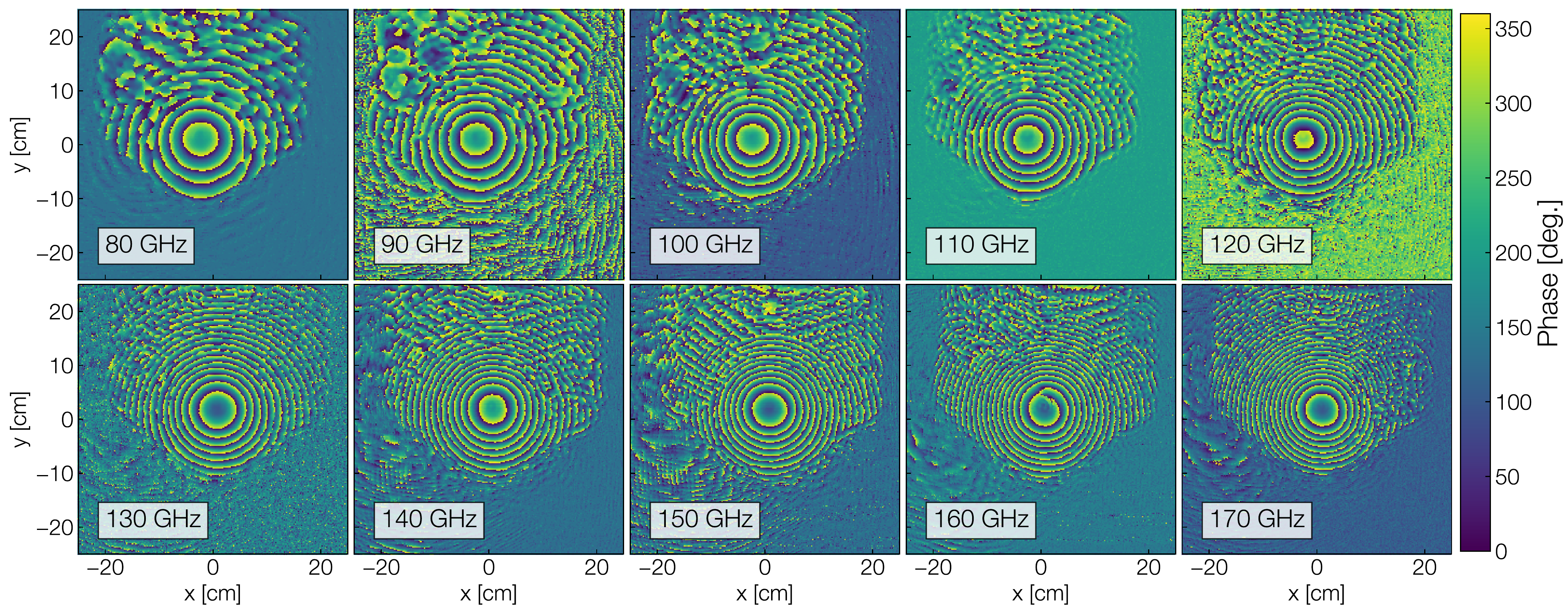}
    \caption{Holography beam map measurements in the mid-frequency band.  Top: Power at each measured frequency (peak-normalized in each map). The hexagonal side-lobes in the power beam maps are higher in power than the signal-to-noise of the system (45\,dB).  Bottom: Wrapped phase at each measured frequency.}
    \label{fig:beam_measurements_all}
\end{figure*}
\section{Measurement Approach}
\label{sec:meas_method}
Here, we describe the hardware and software used in these holography measurements.  Further details can be found in Appendix~\ref{sec:appendix_hardware}.  For this discussion we divide the system into a cryogenic system mounted in the optics tube and a holography system comprised of a source, correlation receiver, and motion system.
\begin{figure*}[ht]
    \centering
    \includegraphics[width = .95\textwidth]{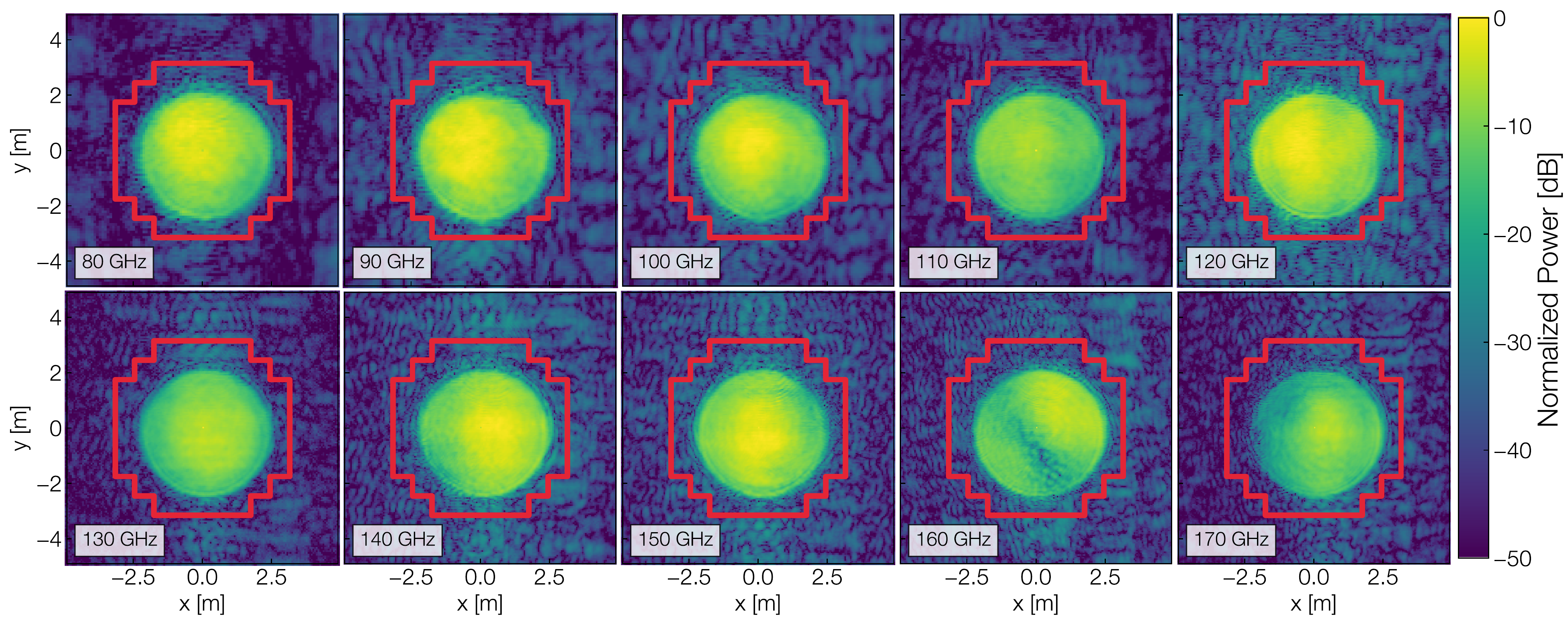}
    \caption{All beam maps propagated at the secondary illumination of the Large Aperture Telescope.  The spilled power to 300\,K is calculated by integrating power outside the boundary of the secondary (red line) with respect to total integrated power of the map.}
    \label{fig:secondary}
\end{figure*}
\subsection{Cryogenic System}
\label{sec:cryo_rec}
The optics tube is housed in a test cryostat called the LATR tester \cite{Harrington_2020}.  The cryostat holds and cools a single optics tube and provides support for detector readout.  This setup supports up to three detector arrays.   In the test configuration, two are bolometric arrays~\cite{2022arXiv220104507H} and the third is used for the holography measurements. 

The holography array consists of a feedhorn array identical to that used for the bolometric detectors, but with standard wave guide flanges at the outputs. A receiver consisting of a round to rectangular wave guide transition and a harmonic mixer is attached to this feed array.  The mixer was designed to operate from 70-110 GHz, but was found to operate satisfactorily up to 170 GHz. For operational simplicity, we used this mixer over our full frequency range from 80-170\,GHz.
A second identical receiver was also connected for redundancy, but not used in the measurements described here.   Two 0-18\,GHz coaxial connections were made from the receiver to connectors at the cryostat wall.  There are heat sinks spaced along the coaxial cables between the focal plane, which was operated at 4\,K while doing holography, and the 300\,K cryostat wall.  A separate cool down was used to measure the loss along these coaxial feed lines.  The loss  was found to be 23\,dB at the LO frequency (10-13\,GHz).  Accurate knowledge of the loss along the feed lines is critical for providing the correct amount of power to the mixers in the focal plane.  The loss at the intermediate frequencies (IF) (100\,MHz) is significantly lower and not critical to the function of this system.
\subsection{Holography System}
\label{sec:meas_hardware}
Figure~\ref{fig:setup} shows a schematic overview of the LATR tester (LATRt) holography hardware.  Two millimeter-wave sources are used to measure the full SO Mid-Frequency (MF) band: F90 (80-120\,GHz) and F150 (130-170\,GHz).  Only one is mounted at a given time (one of which has a multiplication factor of 8, and the other with 12).  These sources are broadcast into the receiver using standard gain feed horns held close to the window (4.5\,cm in F90 and 11.5\,cm in F150 due to the different attenuator waveguide lengths).  We therefore expect different measured beam sizes between the two sets of frequencies, since the beam expands as it leaves the optics tube.
 
A motorized two-dimensional  stage holds the source and is mounted on a support structure above the LATRt.  During a measurement, the source (frequency is fixed) is moved over a $50\times50$\,cm range with 0.25\,cm steps (black arrows in Figure~\ref{fig:setup}).   One map takes roughly 12 hours to complete.

A common local oscillator (LO2 in Figure~\ref{fig:setup}) is fed to two harmonic mixers: 1) picked off from the source and 2) at the output of the cryogenic receiver.  The IF signal from both mixers in the 0-125\,MHz band is amplified and passed to a digital correlator (Casper ROACH2 ~\cite{roach2}) which computes the complex correlation between the two signals~\cite{ches18}.  The FPGA on the ROACH2 board outputs the amplitude and the phase of the correlated output, subdivided into a number of 100\,kHz wide bins.  Only the bin associated with the IF frequency is used in subsequent analysis.  The software to program and analyze output from the FPGA is made public on the \textit{McMahonCosmologyGroup} GitHub page in a package called \verb|holog-exp|~\cite{holog-exp}.  Appendix~\ref{sec:appendix_hardware} provides further details to the hardware of the holography setup.
\section{Results and Interpretation}
\label{sec:results}
\subsection{Near-Field Beam Maps}
Figure~\ref{fig:beam_measurements_all} shows the power and phase of the beam maps at each frequency for which the measurement was carried out.   A variable attenuator at the output of the source was used to optimize the amount of signal entering the optics tube, to ensure power was not too high such that the measurement was saturated, but to also ensure the signal was high enough for signal-to-noise greater than 45\,dB.  As stated in Section ~\ref{sec:meas_method}.\ref{sec:meas_hardware}, the F90 source hangs closer to the window than the F150 source (due to different attenuator lengths), and for this reason, we expect the F90 beams to be smaller than the F150 beams.

The shape of the main beam was found to be in good agreement with simulations at all frequencies.  The asymmetric feature seen in the main beam at 160 GHz is believed to be associated with an extra mode in the round wave-guide of the receiver~\footnote{Impedance of the relevant mode is changing rapidly above its cutoff frequency, leading to a high coupling at 160GHz that falls off as you go higher.  Modeling this n detail requires knowledge of exact mixer coupling geometry, among other experiment details which are not fully known. We have seen similar behavior in other systems.  Therefor, assertion of higher modes explaining 160GHz is based on behavior of similar systems and empirically driven by what we see in the data.
}.  Even with this feature, the radial profile is in very good agreement with the theoretical prediction (within 10\% at the -20dB level).   The hexagonal side-lobe seen in each beam map are associated with scattering from within the optics tube and out the hexagonal window.   The phase indicates the direction of propagation of the beam in space.  This suggests that the field of the near-field side-lobes is diverging (or propagating away from the center of the optical path) less quickly than that of the main beam.  
\begin{figure}[ht!]
    \centering
    \includegraphics[width = .47\textwidth]{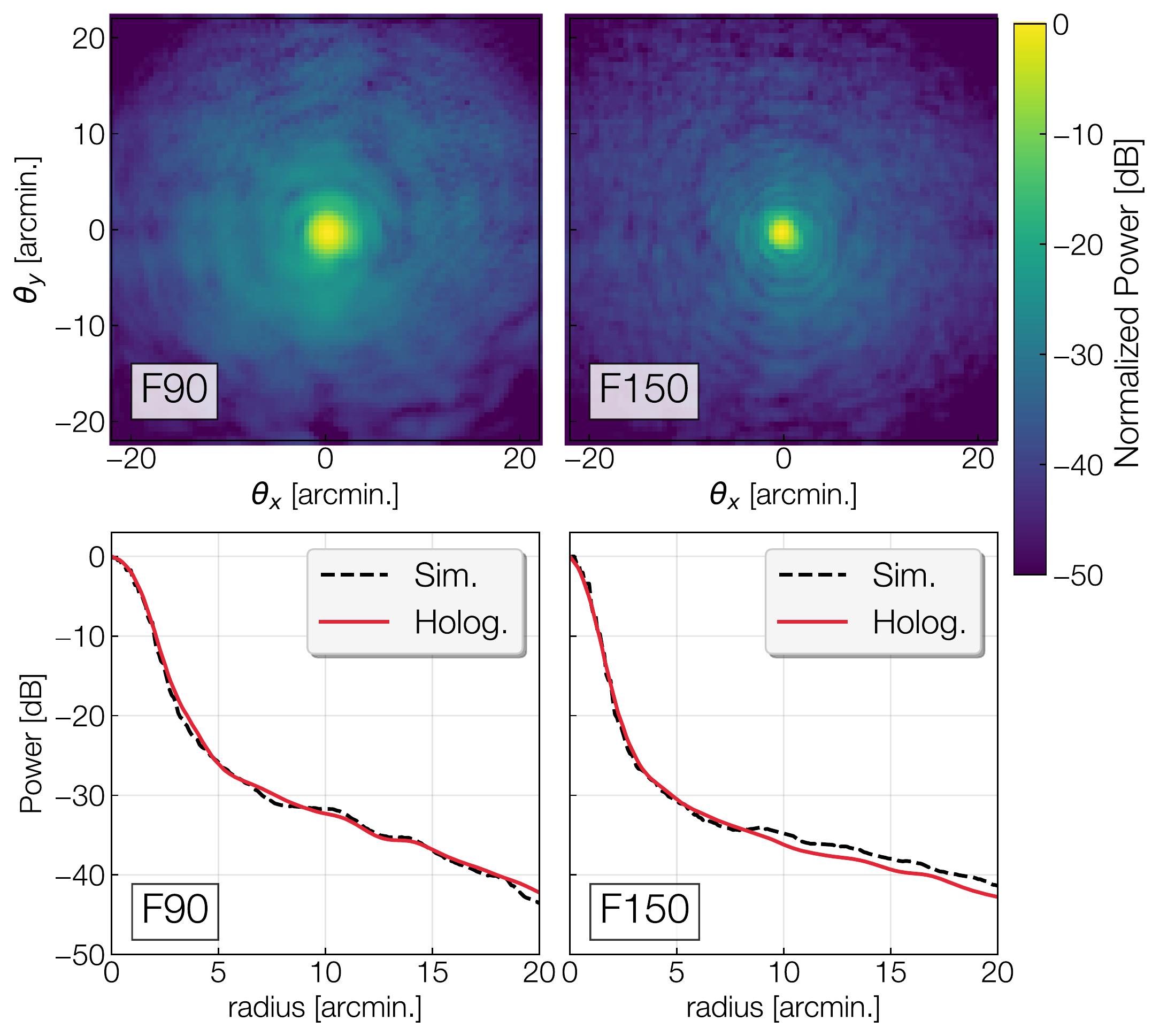}
    \caption{Top row: Using FD, the LATRt measurements are propagated through the LAT from the near-focal plane, to the far-field.  Bottom row: Radial profile of the measured(simulated) far-field beams in the F90 and F150 bands, plotted in red(black).  The lack of diffraction rings in the radial average is explained by the convolution of the measurement(simulation) due to the rectangular source aperture physically in the setup (artificially in the model, see Appendix~\ref{sec:forward_model}).}
    \label{fig:farfields}
\end{figure}
\subsection{Propagation of Fields}
\label{sec:prop_fields}
The performance of the LAT optical system is assessed in detail by using the amplitude and phase of the measured beams to calculate the fields as they propagate through a virtual telescope.  This enables calculation of the far-field beam of the telescope, and the amount of signal "spilled", or lost, at the LAT secondary mirror.  This represents a unique capability of holography measurements which is critical in assessing the overall performance of this system. 
\subsubsection{Fields at Secondary Illumination}
To determine the amount of power spilled to 300\,K, we propagate the measured fields forward and onto the plane of the secondary mirror (approximately 12\,m away from the measurement plane.  This is carried out by using the Fourier relationship between the near-field $E(x,y)$ and far-field $B(\theta_x,\theta_y)$ beams \cite{McIntosh2016,alma_holog}:
\begin{equation}
    B(\theta_x,\theta_y) = \int_{\text{aperture}} E(x,y) e^{ i \frac{2\pi}{\lambda} (\theta_x x + \theta_y y )} dx dy 
    \label{eq:fft}
\end{equation}
where the complex electric field $E(x,y)$ is integrated over the area of the aperture, and $\lambda$ is the wavelength.
\begin{figure*}[ht]
    \centering
    \includegraphics[width = .95\textwidth]{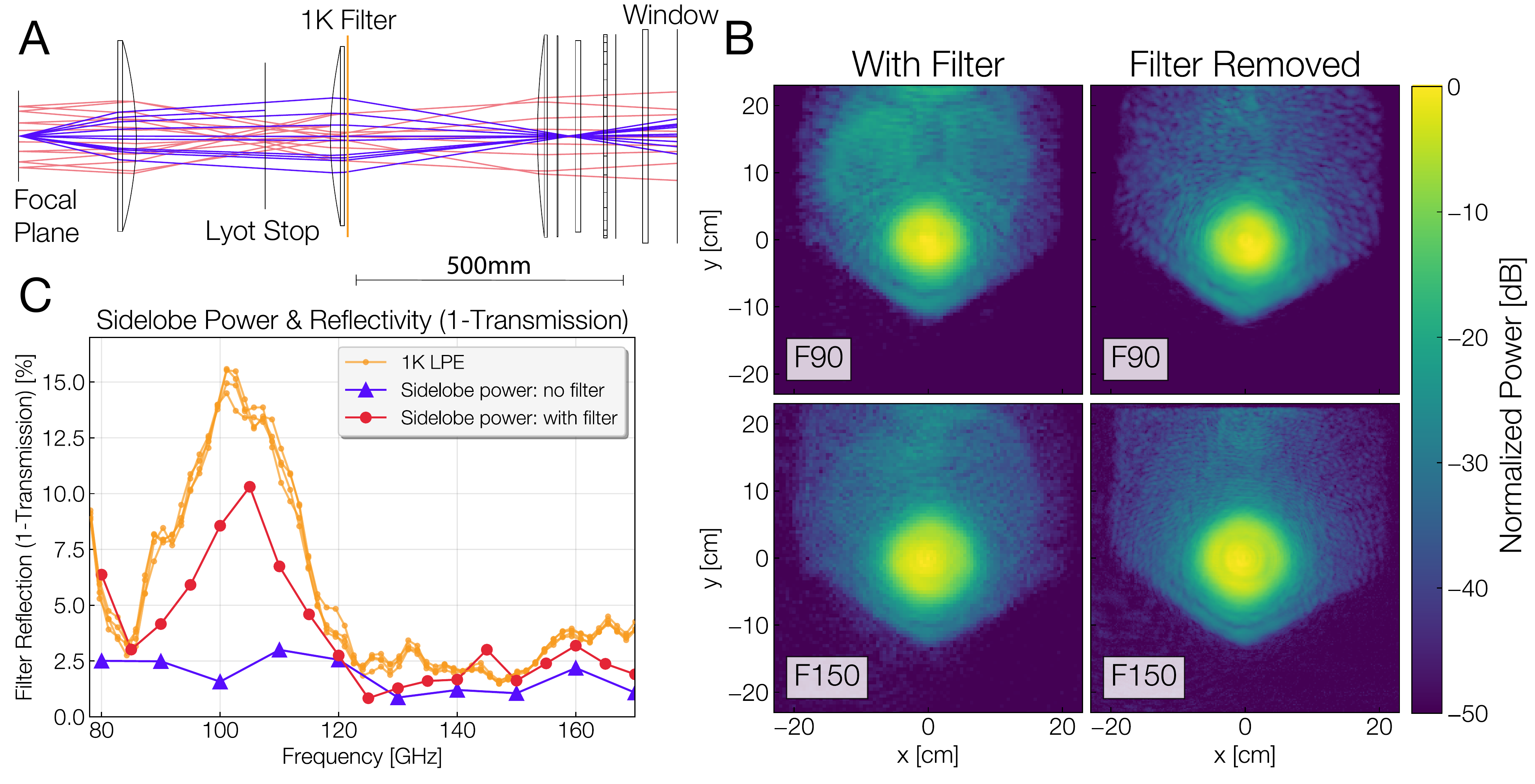}
    \caption{A) Time-reversed ray-trace of the SO LAT Optics Tube (OT) with one filter (orange), near the second lens.  With 0\% on-axis reflectivity of the filter, rays(blue) emerge from the focal plane and exit the OT window.  With an on-axis reflectivity of 5.64\,\% (average LPE filter reflectivity in the full band), rays reflected off of the filter(red) go back to the reflective focal plane (copper and reflective), and then propagate out the front of the window.  This is verified in the simulation, as the main beam in front of the OT shows the shape of the focal plane at -25\,dB.  \,\,B) In orange is the measured reflectivity (1-transmission) of the 6.8\,cm$^{-1}$ LPE filter, which is placed at the 1-K stage in the optics tube.  The red(blue) line shows the measured integrated fractional power outside the main beam at each frequency with(without) the filter in the optics tube.  \,\,C) Band-averaged near-field beam maps with and without the filter in the optics tube.  With the filter, beam maps show extra scattering in the top parts of the map, through the upper portion of the LATRt hexagonal window (hexagonal power around the main beam at roughly -20\,dB due to reflection).}
    \label{fig:filter_info}
\end{figure*}

Figure~\ref{fig:secondary} shows beam maps propagated to the secondary mirror of the LAT, with the boundary  of the secondary mirror in red.  To quantify the spilled power, we integrate the power outside the boundary, and then normalize to the total integrated power of the beam map.  We find the average spilled power in the F90(150) detector band is 0.65 (0.68)\% with no significant frequency dependence.  This is below the design target of 1\% and indicates the sensitivity of SO should not be compromised by spilled power to 300\,K.  This estimate excludes the LATR forecone, a reflective cone of $31.5^\circ$ in front of the LATR focal plane, which reflects spilled power onto the secondary mirror~\cite{2021RNAAS...5..100X}.  Therefor, this method over-estimates the amount of spilled power to 300\,K~\cite{Gudmundsson:21}.

\subsubsection{Far-Fields}
To propagate the measured near-fields into the far-field, we use a virtual telescope to produce the fields from a distant (100\,km) point source on the measurement plane.  We then multiply this field with the measured near-field beam in the corresponding plane.  Integrating the resulting field over the area of the focal plane produces the amplitude and phase of the far-field at that point.  We then rotate the telescope in azimuth and elevation and repeat this process to produce a full beam map in the sky.

The holography source emits its signal out of a rectangular feedhorn with a finite size.  The resulting near-field beam is convolved by this rectangular aperture~\cite{Goodman2005-ne}.  This method of convolution is commonly used when analyzing fields measured with rectangular horn faces~\cite{1141856,Karimipour2019ShapingEW}.  Interpreting this measurement requires accounting for this effect, which amounts to a convolution of the electromagnetic field from the optics tube with the field pattern on the aperture of the feedhorn.  The impact of this convolution is to broaden the F90(150) far-field beam by 12.2(4.7)\% and to create square diffraction spikes in the raw far-field calculation~\cite{Goodman2005-ne}.

We account for this effect with forward modeling, which is described in Appendix~\ref{sec:forward_model}.  To fully simulate the far-field beam of the LAT including the optics tube, we first simulate the optics tube using \verb|solat-optics| as described in~\cite{holog_sim_model}.  This produces the near-field beam at the front of the optics tube, which is then propagated into the far-field the same way the measured near-field beams are propagated through the virtual telescope.  The resulting far-fields after diffraction spike removal are shown in Figure~\ref{fig:farfields}.  These plots are band averaged, including data from 80-110\,GHz in the F90 band and 130-170\,GHz in the F150 band.  The radial binned far-field holography data are compared to simulations.  These comparisons show that the holography data are consistent with the predicted F90(150) FWHM is 2.18(1.38) arcmin and with low ellipticity with no unexpected features such as the "little-buddies" seen in the ACTPol experiment \cite{2021arXiv211212226L,Gudmundsson:21}.
 \begin{figure*}[t!]
    \centering
    \includegraphics[width = .8\textwidth]{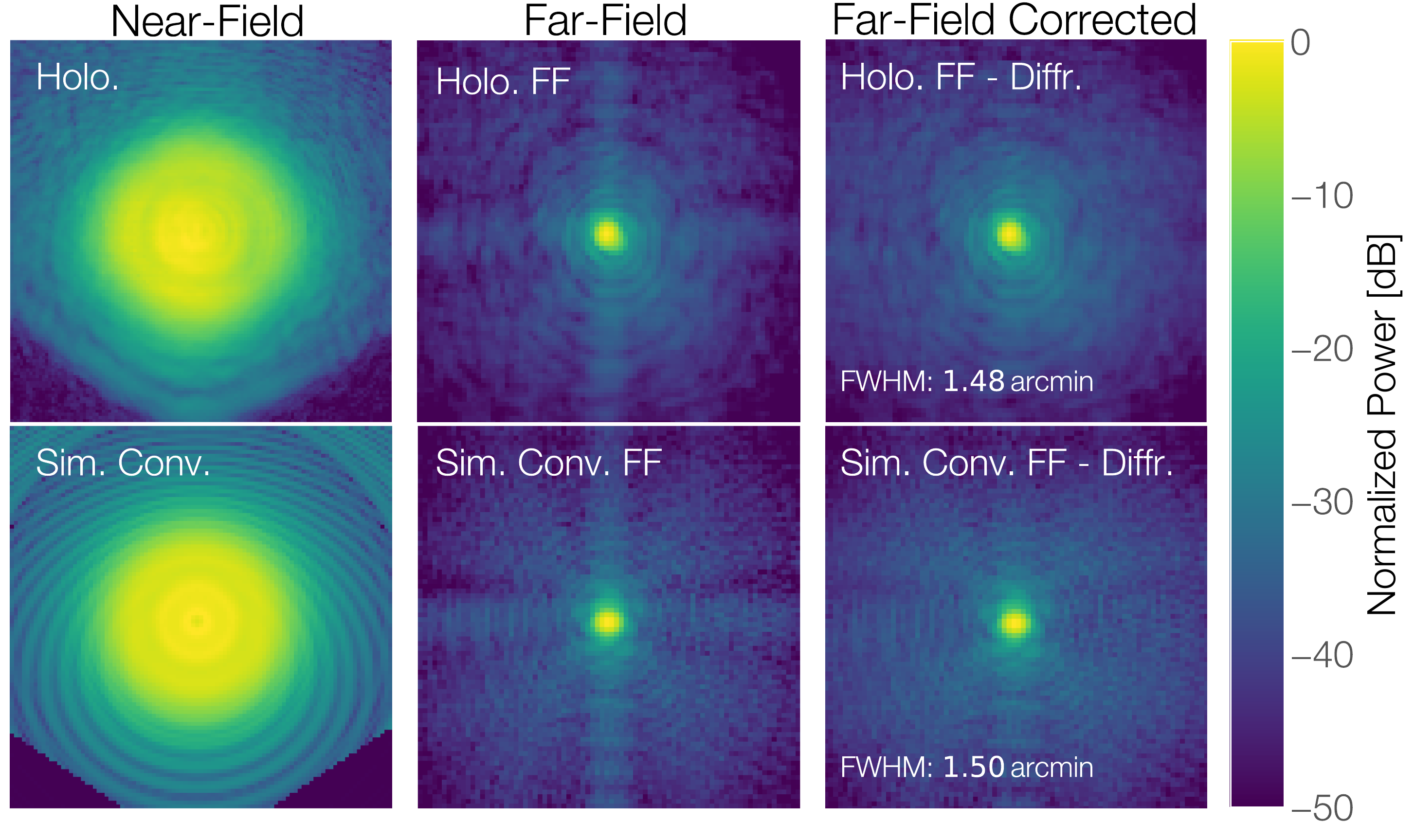}
    \caption{Forward modelling method with the F150 band-averaged holography data. Column 1:  Near-field holography data(top) and convolved simulation with a scattering term(bottom).  Square is $12\times12\,$cm.  Column 2: Far-field holography data(top) and convolved simulation with a scattering term(bottom).  Diffraction spikes consistent with a convolution from a square aperture are present in both the measured far-fields and the simulated far-fields, due to the source horn having a rectangular aperture face.  Square is $20\times20\,$arcmin.  Column 3:  Far-field holography data with diffraction model removal(top) and convolved simulation with a scattering term with diffraction model removal(bottom). Square is $20\times20\,$arcmin.}
    \label{fig:forward_model}
\end{figure*} 
\section{Filter Removal}
\label{sec:filter}

We have described the holography results from the final instrument configuration.  However, in the initial SO configuration, which contained an additional filter (1K Low-Pass Edge (LPE) filter in Figure~\ref{fig:latrt}), we found extra signal outside the main beam from the window (hexagon at ~-20dB) at all frequencies.  While these side-lobes did not significantly change the spilled power to 300\,K, they would have reduced optical efficiency and led to an enhancement of near side-lobes of the on-sky beam.

To study the frequency dependence of side-lobe power, we computed the integrated fractional power outside the main beam.  We define the main beam radius as $13.5$\,cm, where the beam drops below -20\,dB.   Figure~\ref{fig:filter_info}B shows the integrated fractional power outside the main beam as a function of frequency, and also the measured reflectivity of the 1\,K 6.8\,cm$^{-1}$ LPE filter~\cite{10.1117/12.673162}.

Comparing the side-lobe power to the reflectivity measurements of all filters in the optics tube, we noticed the closest resemblance to the 1\,K LPE filter.  To investigate the effect of a reflective 1\,K LPE filter, a simulation in Zemax~\cite{zeemax} shows the expected measured signal due to reflectivity of a filter (Fig. ~\ref{fig:filter_info}A).  The simulation predicts that rays are reflected off the filter and end up outside the main beam as the rays exit the LATRt window.

After removing the LPE filter from the optics tube and repeating the holography measurements, we measured a decrease in side-lobe power.  Figure~\ref{fig:filter_info}C shows the reduced near-field side-lobe power following the removal of the filter, and the comparison of the side-lobe power with the filter in place.  With the filter removed, the new side-lobe fractional power across the band is 1.9\%, a factor of 3 lower.   The tests presented above do not include this filter.  This provides a concrete example of how holography measurements can be used to optimize cryogenic optical systems.
\section{Public Code}
\label{sec:code}
Here, we describe the software used for the holography data acquisition and analysis, all of which are made public.  The software is two-fold; 1) the \verb|solat-optics| module for simulating the near- and far-fields of the LAT telescope from the LATRt holography measurements and 2) Open Source Holography: a website detailing software and hardware to replicate the holography measurements.
\subsection{Optics Simulation}
The \verb|solat-optics| code includes several modules: beam simulation of the LATR optics tube, propagation of measured fields to the secondary mirror, and to the far-field (dependencies:~\cite{2020NumPy-Array,2020SciPy-NMeth,hunter2007matplotlib,reback2020pandas,mpiPython,tqdm}).  This beam simulation includes the lens geometry and computes the beam in the near-field.  The code can be adapted to produce the beam as a function of angle, as was used in this paper, or as a function of position in the measurement plane.  

The propagation analysis code inverts complex beams either from simulations or holography data, corrects for near-field aberrations using ray tracing and returns the complex fields propagated to a desired plane, either near- or far-field.   Notebooks are provided to show how to compute a beam simulation, how to analyze a LATRt holography measurement, and how to propagate a measurement to a desired region.  We invite users to adapt this code to any applications they see fit, but ask that publications using this code cite this paper and that code derived from this work remain public.
\subsection{Open Source Holography}
We also provide an open-source holography GitHub website:  \verb|holog-exp|~\cite{holog-exp}.  The repository provides both hardware and software details for recreating the holography measurements in this paper, and details on how to adapt the setup for future experiments.  The scripts demonstrate how to correlate signals with the FPGA, program the synthesizers, and program the XY stages to produce a holography beam map.  Further details can be found in Appendix~\ref{sec:appendix_hardware}.
\begin{figure*}[t]
    \centering
    \includegraphics[width = .95\textwidth]{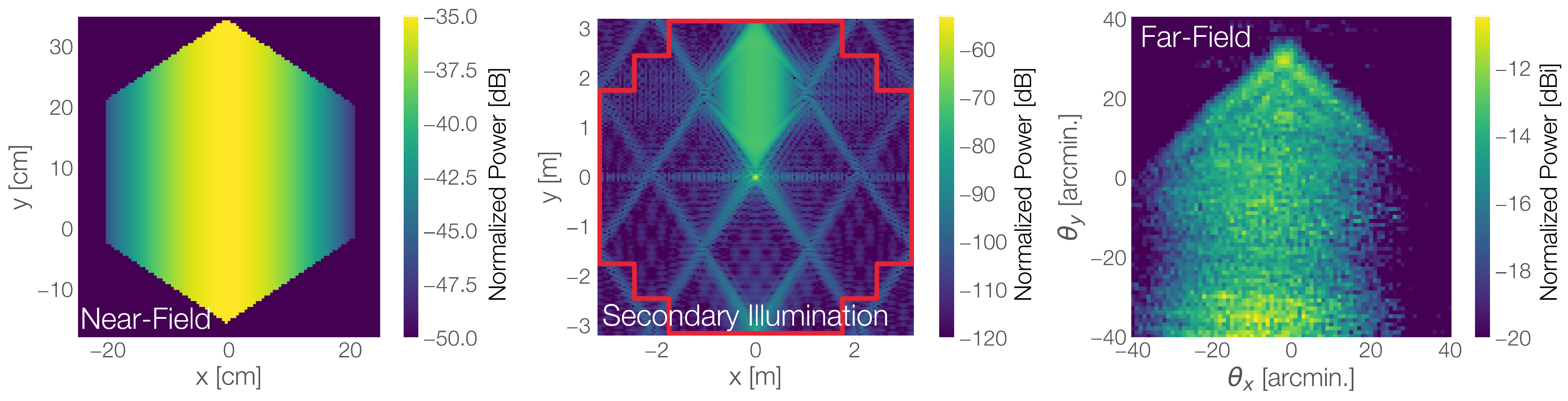}
    \caption{Simulated scattering term (power) in the near-field(left), at the secondary illumination(middle) and propagated to the far-field(right).  Each subplot is the band-averaged simulation over all F150 frequencies.}
    \label{fig:scattering_forward}
\end{figure*}
\begin{figure}[ht]
    \centering
    \includegraphics[width = .47\textwidth]{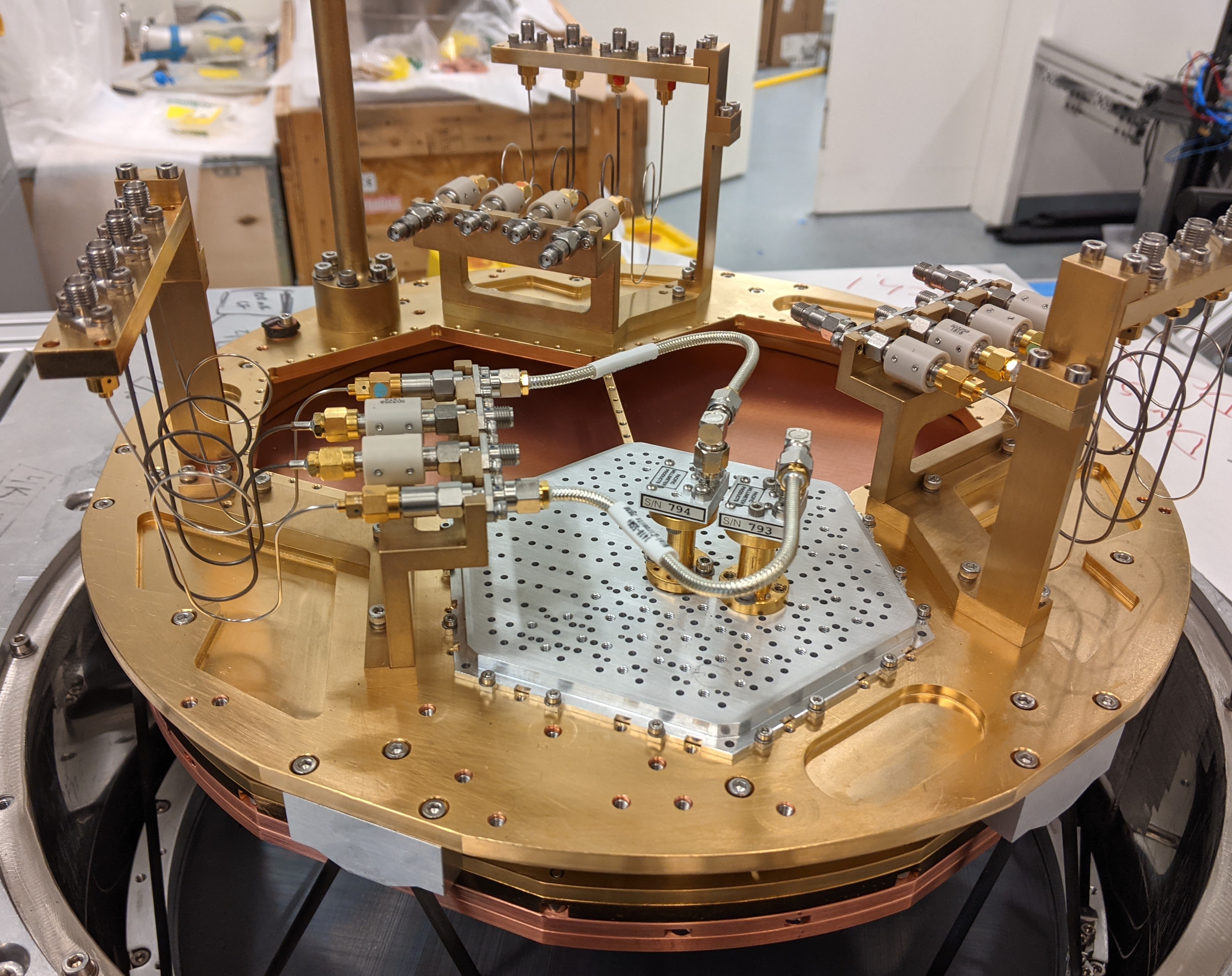}
    \caption{The Simons Observatory Large Aperture Telescope optics tube focal plane readout, which is cooled to 4\,K during measurements.  The holography receivers (two receivers for redundancy) are approximately 7.4\,cm from the center of the focal plane.}
    \label{fig:fpa}
\end{figure}
\section{Conclusion}
\label{sec:discussion}
Refractive holography enables the testing of optical performance prior to deployment, and propagating the measurements into the far-field to predict the beam of the telescope.  We have presented holography measurements of the SO LAT optics tube and analysis methods determining its optical performance, including open-source holography for repeating and adapting the experiment.  We further provide an open-source package for simulating near-field holography measurements and propagating the measurements into the far-field using FD. 

From these data, we characterize the optical performance of the LATR optics tube.  We compare near- and far-field measurements to simulations.  After propagating the beams to the plane of the LAT secondary mirror, we find sub-percent power spilled to 300\,K.  We further find the far-field measured beams to be 2.18(1.38) arcmin FWHM in the F90(F150) band.

We provide three open-source software packages.  The first developed for this work, (\verb|solat-optics|~\cite{holog_sim_model}), models the LATRt holography measurements and is customizable to include arbitrary optics and adaptable for other optics experiments.  The second, (\verb|holog-exp|~\cite{holog-exp}), details the hardware and data acquisition software required in this experiment.  And lastly, we publish the data and Python scripts for recreating all figures in this paper~\cite{knowledge}.

The approach demonstrated here is broadly applicable to the characterization of millimeter-wave optical systems.  The ability to characterize the optical performance and systematics of the optics tube allowed us to determine one source of spurious reflections, and avoid systematics during future observations.  
\section{Backmatter}
\begin{backmatter}
\bmsection{Funding}
This work was funded by the Simons Foundation (Award \#457687, B.K.).  G.E. Chesmore is supported by the National Science Foundation Graduate Student Research Fellowship (Award \#DGE1746045).  Z. Xu is supported by the Gordon and Betty Moore Foundation through grant GBMF5215 to the Massachusetts Institute of Technology.  G. Coppi is supported by the European Research Council under the Marie Sk\l{}odowska Curie actions through the Individual European Fellowship No. 892174 PROTOCALC.  M.D. Niemack acknowledges support from NSF grant AST-2117631.

Work supported by the Fermi National Accelerator Laboratory, managed and operated by Fermi Research Alliance, LLC under Contract No. DE-AC02-07CH11359 with the U.S. Department of Energy. The U.S. Government retains and the publisher, by accepting the article for publication, acknowledges that the U.S. Government retains a non-exclusive, paid-up, irrevocable, world-wide license to publish or reproduce the published form of this manuscript, or allow others to do so, for U.S. Government purposes.
\bmsection{Disclosures} The authors declare no conflicts of interest.
\bmsection{Data Availability}
Data underlying the results presented in this paper are made publicly available at knowledge.uchicago.edu~\cite{knowledge}, along with a Jupyter Notebook which recreates all data figures presented.
\bmsection{Author Contributions}
G.E.C. and J.M. devised the holography experiment hardware and readout.  K.H. led the LATRt optical testing program.  P.A.G. provided the Zemax ray-trace in Figure~\ref{fig:filter_info}.  G.E.C. wrote analysis and simulation code.  G.E.C and J.M. wrote the manuscript with input from all authors.  J.M. supervised the research.
\end{backmatter}
\appendix
\section{Open-Source Holography}
\label{sec:appendix_hardware}
Figure~\ref{fig:setup} shows a schematic of the RF electronics.  Two local oscillators (LOs) produce signals between 10 and 13 GHz. The LO1 synthesizer produces a signal between 10 and 13\,GHz, while LO2 produces the same frequency with some offset $f_{\text{offset}}$ (this offset is chosen to be 10\,MHz).  The offset frequency is what will eventually produce an intermediate frequency exiting the mixer diplexers.  The purpose of the mixer diplexers is to ensure the signal from the first LO travels to the two mixers, and then ensures that the IF output of the mixers travels in the opposing direction down the RF chain to the FPGA.

The LO1 signal goes to the active multiplying chain, where it is multiplied by 8(12), obtaining frequencies in the F90(150)-band.  Prior to leaving the source horn, -10dB of the signal splits off, and mixes with LO2 in the harmonic mixer, producing an intermediate frequency IF$_1$ which then goes through one of the mixer diplexers and to the FPGA.  The rest of the signal leaves the source horn, through the components of the LATR optics, and the signal reaching the back of the optics tube mixes with LO2 in a GaAs harmonic mixer, and the subsequent intermediate frequency, IF$_2$, which also travels to the FPGA.  The FPGA used for these measurements is the Re-configurable Open Architecture Computing Hardware (ROACH-2) board, which correlates the reference and modulated signals~\cite{roach2}.

The signal from the source(receiver) needs to be amplified due to high loss levels in the coax path from the harmonic mixer on the source to the reference mixer-diplexer (LATRt readout chain).  To overcome this loss, amplifiers with attenuators  increase the signal in the RF chain prior to entering the mixer diplexer.  The setup uses low phase variation coaxial cables leftover from the DASI experiment~\cite{CHURCH20031083}.  The phase repeatability of the holography setup is within $\approx3^{\circ}$.  We further note that any drift in the map would present itself in the phase map.  This phase drift would be removed during the propagation into the far-field when we optimize the position of the beam map in the LAT focal plane. 

The source moves in a 2-D grid above the LATRt with motorized XY stages~\cite{stages}.  The source mounts to the stages such that the signal points downwards towards the LATRt window.  The laboratory is over 7.5\,m tall, and therefore we expect reflection from the walls to be diffuse.  Therefore, the reflected signal is diluted before reflecting into the testing system.  The dominant reflections are from reflections within the optics tube, as the hexagon side-lobe is the dominant side-lobe feature in the beam maps (Figure~\ref{fig:beam_measurements_all}).

Figure~\ref{fig:fpa} shows the holography receiver readout at the back of the optics tube focal plane.  An SO MF feedhorn array is adapted for the holography experiment.  On the readout side of the array, attachment screws are added for attaching a circular to rectangular transition waveguide.  The transition waveguide connects the back of the focal plane (circular) to the GaAs W-band harmonic mixer (rectangular).  Though the design of the W-band harmonic mixer is optimized for F90 frequency readout, the W-band harmonic mixer is used for both F90 and F150 measurements (the entire SO MF band), allowing for a wider band of measurements without separate LATRt cool-downs
\begin{table}[ht]
\centering
\begin{tabular}{|c|c|c|c|c|}
\hline
F\,[GHz] & \multicolumn{3}{c|}{S/N [dB]}\\
& NF& Sec.& FF\\
\hline
 90      
         & 20& 21.7 & 40.0 \\
         & 40& 37.0 & 43.9 \\
         & 60& 45.4 & 43.9 \\
 150     
         & 20& 18.0 & 39.3 \\
         & 40& 35.5 & 43.8 \\
         & 60& 46.0 & 43.8 \\
 \hline
\end{tabular}
\caption{Near-field simulated measurement signal-to-noise and resulting simulated side-lobe power (at the secondary illumination and into the far-field).  Signal-to-noise is calculated as the standard deviation of the signal outside an 8.5\,cm radius of the peak-normalized beam (same resolution as Table~\ref{tab:fft_grid}).}
\label{tab:fft_sn}
\end{table}
\section{Measurement Requirements}
\label{sec:err_prop}
Here, we discuss the measurement requirements to meet specific far-field grid resolution and range from near-field data.  Near-field beams with three signal-to-noise levels are propagated into the far-field (for 90\, and 150\,GHz near-field beams); we simulate the near-fields to have 20, 40, and 60\,dB signal-to-noise.  The signal-to-noise propagated to the secondary illumination and far-field is listed in Table~\ref{tab:fft_sn}. 

When planning the near-field scan, we consider the resolution and how the near-field grid propagates into the plane of the secondary, due to the Fourier relationship between near- and far-fields (Eq.~\ref{eq:fft}), and into the telescope's far-field through the modeling described in Section~\ref{sec:prop_fields}. Table~\ref{tab:fft_grid} summarizes the scan size and resolutions and resulting far-field size and resolution grids used in the holography measurements presented here.
\begin{table}[htb]
\centering
\begin{tabular}{|c|c|c|c|c|c|c|}
\hline
F\,[GHz] & \multicolumn{2}{c|}{NF [cm]}&\multicolumn{2}{c|}{Sec. [m]} & \multicolumn{2}{c|}{FF [arcmin]} \\
 & Size & Res. & Size & Res.& Size & Res.\\ \hline
 90 & 50& 0.25 & 9.52& 0.13& 119.72 & 0.60\\
 150 & 50& 0.20 &6.66 &0.06 &128.40&0.52\\
 \hline
\end{tabular}
\caption{Near-field scan size and resolution, and resulting scan size and resolution at the secondary illumination and in the far-field.}
\label{tab:fft_grid}
\end{table}
\section{Forward Modelling}
\label{sec:forward_model}
As introduced in Section~\ref{sec:results}.~\ref{sec:prop_fields}, the holography source emits from a rectangular feedhorn, and therefore result in a convolution of the electromagnetic field from the optics tube with the field pattern on the feedhorn aperture.  Convolving the simulated fields increases the F90(150) far-field beam by 12.2(4.7)\% and results in horizontal and vertical diffraction spikes in the raw far-field calculation~\cite{Goodman2005-ne}.

We carry out the forward modelling by building a simulation of the optics tube beam pattern at the measurement plane.  This model includes an empirical model of the scattering of the optics tube with the hexagonal outline to model the boundary of the optics tube window, and a similar amplitude and phase to what is measured.  The resulting simulated F90(150) beam matched the measured FWHP beam width within in 1.73(0.7)\%.  The simulation also had the horizontal and vertical diffraction spikes in the far-field due to the impact of the convolution with the holography source feed pattern~\cite{Goodman2005-ne}.  The full forward modelling process is shown in Figure~\ref{fig:forward_model}.

For visualization purposes, these spikes are removed by subtracting a model $D(\theta_x,\theta_y)$ (amplitude of the electric field) consisting of a $\sinc$ function with a Gaussian width along its narrow direction equal to the beam width (Eq.~\ref{eq:model_conv}), where $\theta_x$ and $\theta_y$ are elevation and azimuth, and we fit the 2 parameters $\sigma_{\theta_x}$ and $\sigma_{\theta_y}$.  This model is based on the predicted Fraunhofer diffraction pattern from a rectangular aperture~\cite{Goodman2005-ne} (e.g., the feed) and was shown to match the simulations.
\begin{equation}
    D(\theta_x,\theta_y) = \exp^{-(\theta_x^2/4\sigma_{\theta_x}^2 +\theta_y^2/4\sigma_{\theta_y}^2 )}\sinc{\theta_x}\sinc{\theta_y}
    \label{eq:model_conv}
\end{equation}
The holography measurements showed scattering from within the optics tube (hexagonal shape at ~-20\,dB in Figure~\ref{fig:beam_measurements_all}).  To understand how this scattering propagates through the telescope, we add a scattering term (with both amplitude and phase) to the simulated beam, and then propagate this beam into the far-field.  We also investigate how the side-lobes measured outside the main beam (Fig.~\ref{fig:beam_measurements_all}) propagate into the far-field (Fig.~\ref{fig:scattering_forward}).  Reflections are known to be a problem in near-field beam mapping~\cite{2020JLTP..199..156Y,7740846,387181}.  However, the inferred amplitude of the probe is small, and we do not correct for reflections.  The side-lobe spreads out and is localized to 2 meters from the center of the primary and secondary mirrors, and then leads to a $0.85^{\circ}$ diffuse structure on the sky that is at the $\approx -15$\,dBi level.
\bibliography{sample}
\end{document}